\documentstyle[prl,aps,twocolumn]{revtex}
\begin{document}
\draft
\title{Microscopic dynamics underlying the anomalous diffusion}
\author{G. Kaniadakis$^1$\footnote{e-mail: kaniadakis@polito.it}, G. Lapenta$^2$\footnote{e-mail: lapenta@polito.it}}
\address{ Istituto Nazionale per la Fisica della Materia - Unit\`a del
 Politecnico di Torino\\ $^1$Dipartimento di Fisica and $^2$Dipartimento di Energetica -
 Politecnico di
 Torino\\
Corso Duca degli Abruzzi 24, 10129 Torino, Italy}
\date{\today}
\maketitle

\begin {abstract}
The time dependent Tsallis statistical distribution describing
anomalous diffusion is usually obtained in the literature as the
solution of a non-linear Fokker-Planck (FP) equation [A.R.
Plastino and A. Plastino,   Physica A, {\bf 222}, 347 (1995)]. The
scope of the present paper is twofold. Firstly we show that this
distribution can be obtained also as solution of the non-linear
porous media equation. Secondly we prove that the time dependent
Tsallis distribution can be obtained also as solution of a linear
FP equation [G. Kaniadakis and P. Quarati, Physica A, {\bf 237},
229 (1997)] with coefficients depending on the velocity, that
describes a generalized Brownian motion. This linear FP equation
is shown to arise from a microscopic dynamics governed by a
standard Langevin equation in presence of multiplicative noise.
\end {abstract}


\section{Introduction}

Recently, the Tsallis thermostatistics has received considerable
attention in light of its growing application to a variety of
physical systems [1]. The research has focused both on fundamental
and phenomenological aspect of the issue [2].

Particular attention has been devoted to the issue of anomalous
diffusion, where a significant experimental evidence has been
gathered (see Ref. [3] for a detailed bibliography). The
description of a diffusive process (either classic or anomalous)
is performed generally by adopting a time dependent formalism. The
Tsallis distribution namely
\begin{eqnarray}
p(v)=\frac{1}{Z_q} \left [ 1- (1-q) \beta v^2 \right ]^{1/(1-q)} \
\ , \nonumber
\end{eqnarray}
with $Z_q=\int_{\cal R} dv \left [ 1- (1-q) \beta v^2 \right
]^{1/(1-q)}$, has been first derived starting from the generalized
entropy
\begin{eqnarray}
S_q=\frac{1}{q-1}\left[1-\int_{\cal R} dv \,\, p^q \right] \ \ ,
\nonumber
\end{eqnarray}
using the maximum entropy principle under the constraint of
conservation of particle number and energy, by solving the
variational problem: $\delta\left[ S_q-\beta E -\alpha N
\right]=0$.

Similarly to the classic Boltzmann distribution, the Tsallis
distribution can be also obtained as the steady-state distribution
of a time dependent Fokker-Planck (FP) equation. Recently, the
research on the derivation of the Tsallis distribution from FP
equations has produced considerable results [3-13]. The research
in this area can be classified in one of two classes.

First, linear FP equations are considered with diffusion and drift
coefficients depending  on the velocity. The dependence is chosen
to lead to the Tsallis distribution as the equilibrium solution of
the FP equation. Within the linear approach, two different choices
of the drift and diffusion coefficients have been proposed.
Stariolo [4] chooses a constant diffusion coefficients and alters
the drift coefficient to include a generalized potential depending
on the Tsallis parameter $q$. This approach introduces a more
relevant modification of the classic Browninan approach. In ref.
[5], instead, the classic Brownian drift coefficient have been
considered, but with a modified diffusion coefficient to include a
quadratic velocity dependence.  The two linear approaches
described above are in reality just examples of an infinite class
of linear FP models that give a Tsallis equilibrium distribution
[6]. Clearly, the selection of a specific linear model among the
class requires the introduction of other criteria beyond the
simple requirement of leading to an equilibrium Tsallis
distribution.

Second, non-linear FP have been shown to lead to equilibrium
Tsallis distributions. This approach, introduced by Plastino and
Plastino [7] and continued by various authors [3,8-13], introduces
a diffusion coefficient depending on powers of the distribution
function. The drift, instead, can be equal to zero or described as
in the classic Brownian motion. This latter approach, besides its
elegance and simplicity, admits time dependent solutions
characterized by retaining at every time the form of a Tsallis
distribution. This self-similarity of the evolution represent an
important property of the non-linear approach.

The present paper deals with the question of whether the linear
[5] and non-linear [7] FP approach to the derivation of the
Tsallis distributions are equivalent. The answer proved here is
that indeed the two approaches are equivalent, in the sense that
they both allow the presence of self-similar transients where the
system is characterized by the Tsallis distribution at every
instant.

In order to explain the microscopic origin of the anomalous
diffusion associated to the non-linear FP equation of ref. [7]
Borland suggested a feedback from the macroscopic level to the
microscopic one [13]. In the present work, we show that the
non-linear FP equation of ref. [7] but also the well known in the
literature non-linear porous media equation, considered recently
in the frame of Tsallis thermostatistics in ref. [8],  can be
recast in the equivalent linear FP equation of ref. [5]. This
important result allows a deeper interpretation of the non-linear
FP equation describing the anomalous diffusion in terms of a
linear Langevin microdynamics in presence of a multiplicative
noise.

The present work is organized as follows. In Section II, a
Generalized Brownian (GB) motion is derived from the Langevin
equation in presence of multiplicative noise. In Section III, the
GB motion is shown to lead to a macroscopic motion described by
the linear FP equation of ref. [5] that admits as a solution a
class of time dependent Tsallis statistical distributions. In
Section IV and V, the same  distribution are shown to represents
states governed also by the non linear FP equation of ref. [7] and
by the non-linear porous media equation respectively. Finally, in
Section VI conclusions are drawn.

\section{Generalized Brownian Motion}

We consider the microscopic process described by the following
Langevin equation:
\begin{equation}
\frac{d v(t)}{d t}+h(t,v)=g(t,v)\Gamma(t) \ \ ,
\end{equation}
with
\begin{equation}
<\Gamma(t)>=0 \ \ ,
\end{equation}
\begin{equation}
<\Gamma(t)\Gamma(t')>=2 \delta(t-t') \ \ .
\end{equation}
The quantity $- m h(t,v)$ is the deterministic force acting on a
particle of mass $m$ and velocity $v(t)$ while  $m g(t,v) \Gamma
(t)$ is a stochastic force acting on the particle, with $\Gamma
(t)$ a Gaussian random variable with zero mean and delta
correlation function. The presence of $g(t,v) $ in eq.(1) implies
that the particle is subject to a multiplicative noise. The
distinction between additive (when $g(t,v)=const.$) and
multiplicative noise (when $g(t,v)\neq const.$) is very
significant when $g(t,v)$ is a time dependent function. In this
case arise naturally the question related to the definition of the
stochastic integral (Ito or Stratonovich definition). For a more
detailed discussion on multiplicative noise, see ref. [14]. The
microscopic process described by eq. (1) implies a macroscopic
process described by the following linear FP equation:

\begin{eqnarray}
\frac{\partial p(t,v)}{\partial t}=\frac{\partial}{\partial v}
\Bigg\{\left[J(t,v)+ \frac{\partial D(t,v)}{\partial v}\right]
p(t,v) && \nonumber \\ +D(t,v)\frac{\partial p(t,v) }{\partial v}
\Bigg\} && \ \ ,
\end{eqnarray}
where the drift coefficient $J(t,v)$ and the diffusion coefficient
$D(t,v)$ have the following expression
\begin{equation}
J(t,v)=h(t,v) \ \ ,
\end{equation}

\begin{equation}
D(t,v)=g(t,v)^2 \ \ ,
\end{equation}
obtained using the Ito definition for the stochastic integral.
Note that, for Brownian motion:
\begin{equation}
J(t,v)= \gamma(t)v \ \ ,
\end{equation}

\begin{equation}
D(t,v)=c(t) \ \ ,
\end{equation}
the drift current in eq.(4)
\begin{equation}
j_{drift}=\left[J(t,v)+ \frac{\partial D(t,v)}{\partial
v}\right]p(t,v) \ \ ,
\end{equation}
is simplified as:
\begin{equation}
j_{drift}=\gamma(t)v p(t,v) \ \ ,
\end{equation}
and the current velocity $j_{drift}/p$ becomes simply proportional
to the viscous force $-m h(t,v)= - m \gamma(t) v$ of the
microscopic process.

A problem arises in conjunction to the results just obtained.
Whether other motions, besides the Brownian motion, are
characterized by a current velocity proportional to the viscous
force. This issue corresponds to the existence of other solutions
of the following equation for the unknown functions $D(t,v)$ and
$J(t,v)$
\begin{equation}
J(t,v)+ \frac{\partial D(t,v)}{\partial v} =\theta(t)J(t,v) \ \ ,
\end{equation}
in addition to the solution (7),(8), relative to the Brownian
motion. The issue is easily resolved and other solutions can be
founded. The more general solution is formed by copies of
functions $J(t,v)$ and $D(t,v)$ where $D(t,v)$ is given by
\begin{equation}
D(t,v)=c(t)+[\theta(t)-1]\int J(v) dv \ \ ,
\end{equation}
while $J(t,v)$ remains arbitrary. The simplest solution, for which
$J(t,v)$ is given by (7) provides the definition for a new
generalized Brownian (GB) motion [5].

\section{Linear Fokker-Planck equation}

We consider the FP equation (4) for the GB processes. With the
introduction of  the dimensionless time $\tau$:
\begin{equation}
d\tau=\theta(t)\gamma(t)dt \ \ ,
\end{equation}
and the functions $D(\tau)$, $\beta(\tau)$ and parameter $q$:
\begin{equation}
D(\tau)= \frac{c(t)}{\theta(t)\gamma(t)} \ \ ,
\end{equation}
\begin{equation}
(1-q)\beta(\tau)=\frac{1-\theta(t)}{2c(t)} \ \ ,
\end{equation}
the diffusion coefficient (12) with drift coefficient given by (7)
can be written in the following form
\begin{equation}
D(\tau,v)= D(\tau) \left[1-(1-q)\beta(\tau)v^2\right] \ \ ,
\end{equation}
while after taking into account (7), the FP equation (4)  becomes
[5]:
\begin{eqnarray}
\frac{\partial p(\tau,v)}{\partial \tau }=
\frac{\partial}{\partial v}\Bigg\{vp(\tau,v)+ D(\tau) && \nonumber
\\ \times\left[1-(1-q)\beta(\tau)v^2\right] \frac{\partial p(\tau,v)
}{\partial v} \Bigg\} && \ \ .
\end{eqnarray}

The time-dependent solutions of eq.(17) are sought using the
following ansatz:
\begin{equation}
p(\tau,v)=\frac{1}{Z_q(\tau)} \left [ 1- (1-q) \beta(\tau) v^2
\right ]^{1/(1-q)} \ \ .
\end{equation}
The above ansatz requires the solution to conserve at every time
the form of a Tsallis distribution with time dependent parameters
$Z_q$ and $\beta$. The time dependence of the two parameters
determines the actual solution and is obtained easily substituting
ansatz (18) in eq.(17). It follows that the equations determining
the evolution of $Z_q(\tau)$ and $\beta(\tau)$ are identical to
the equations for the Brownian motion:

\begin{equation}
\frac{Z_q(\tau)}{Z_q(0)}=\left [\frac{\beta(0)}{\beta(\tau)}\right
]^{1/2} \ \ ,
\end{equation}
\begin{equation}
\frac{d \beta(\tau)}{d\tau}= 2 \beta(\tau) -4D(\tau)\beta(\tau)^2
\ \ .
\end{equation}
The result above justifies the use of the term Generalized
Brownian motion used to name the process defined by eqs. (7,12).
From eq. (20) the condition below follows
\begin{equation}
2\beta(\infty) D(\infty)=1 \ \ ,
\end{equation}
again in complete similarity with Browinan motion. Eq. (20) is
solved easily with the the substitution $y=\beta^{-1}$ that
linearizes the equation:
\begin{equation}
\beta(\tau)\!=\!\beta(\infty)\left\{\!1\!+\!\left [\displaystyle
\frac{\beta(\infty)}{\beta(0)}\!-\!1\!+\!a(\tau)\!\right]
\exp{(\!-2\tau)}\!\right\}^{\!-1} \!,
\end{equation}
with
\begin{equation}
a(\tau)=2\int_0^\tau \left[\frac{D(\tau)}{D(\infty)}-1\right]
\exp{(-2\tau)}d\tau \ \ .
\end{equation}
From eq. (19) it follows
\begin{equation}
Z_q(\tau)\beta(\tau)^{1/2}=Z_q(0)\beta(0)^{1/2}=N_q  \ \ .
\end{equation}
The constant $N_q$ is determined starting from the expression of
$Z_q(\tau)$ given by
\begin{equation}
Z_q(\tau)=\int_{-\infty}^{+\infty} dv \, \, \left [ 1- (1-q)
\beta(\tau) v^2 \right ]^{1/(1-q)} \ \ .
\end{equation}
For $q\geq 1$ [15], it results:
\begin{equation}
N_q=\frac{q+1}{2}\sqrt{\frac{q-1}{\pi}}\frac{\Gamma[1/2+1/(q-1)]}
{\Gamma[1/(q-1)]} \ \ .
\end{equation}
The final solution of eq. (17) has the form
\begin{equation}
p(\tau,v)=N_q \beta(\tau)^{1/2} \left [ 1- (1-q) \beta(\tau) v^2
\right ]^{1/(1-q)} \ \ ,
\end{equation}
where $\beta(\tau)$ is given by eqs. (22-23).

\section{Non Linear Fokker-Planck equation}

The scope of the present and the next section is to show that the
time dependent solution (27) obtained here, of the linear FP
equation (17) proposed in [5], is also solution of non linear FP
equations which can be obtained from the linear FP (17). The goal
of the present section is to investigate the relationship between
the linear FP (17) and the non-linear FP equation proposed by
Plastino and Plastino [7].

We start the proof by noting that eq. (27) allows us to write
\begin{equation}
1-(1-q)\beta(\tau)v^2=N_q^{q-1}\beta(\tau)^{(q-1)/2}p(\tau,v)^{1-q}
\ \ .
\end{equation}
Besides the following time dependent function is defined:
\begin{equation}
D_1(\tau)=\frac{N_q^{q-1}}{2-q}D(\tau)\beta(\tau)^{(q-1)/2} \ \ .
\end{equation}
Then, it follows that eq.(17) can be rewritten as:
\begin{equation}
\frac{\partial p(\tau,v)}{\partial \tau}=\frac{\partial}{\partial
v} \left\{ v p(\tau,v)+ D_1(\tau) \frac{\partial}{\partial v}
[p(\tau,v)]^{2-q} \right\}  \ \ ,
\end{equation}
that is identical to the equation proposed by Plastino and
Plastino that was solved using the same ansatz (18) used above but
assuming that $D_1(\tau)$ is constant.

The procedure outlined in the previous section leads to the same
relationship (19) between $Z_q(\tau)$ and $\beta(\tau)$ and
$\beta(\tau)$ is governed by the following evolution equation:
\begin{equation}
\frac{d \beta(\tau)}{d\tau}= 2\beta(\tau)
-2\beta(\infty)^{(q-3)/2}\frac{D_1(\tau)}{D_1(\infty)}
\beta(\tau)^{(5-q)/2} .
\end{equation}
The condition (21) transforms now as
\begin{equation}
2\beta(\infty)^{(3-q)/2}D_1(\infty)=
 \frac{N_q^{q-1}}{2-q} \ \ .
\end{equation}

As above, a transformation $y=\beta^{(q-3)/2}$ linearizes eq.(31)
and  the general solution follows easily:
\begin{eqnarray}
\beta(\tau)=\beta(\infty)\Bigg(1+\Bigg \{\left[\displaystyle
\frac{\beta(\infty)}{\beta(0)}\right]^{(3-q)/2}-1 \nonumber \\
+b_q(\tau)\Bigg\} \exp{[(q-3)\tau]}\Bigg)^{2/(q-3)} \ \ ,
\end{eqnarray}
\begin{equation}
b_q(\tau)\!=\!(3\!-\!q)\int_0^{^{\!\textstyle{(3-q)\tau \over
2}}}\!\!\! \left[\!\frac{D_1(\tau)}{D_1(\infty)}\!-\!1\!\right]
\exp{[(q\!-\!3)\tau]}d\tau \ .
\end{equation}

The complete solution of eq. (30) is given by eq. (27) where now
$\beta(\tau)$ is expressed as a function of $D_1(\tau)$ by eqs.
(33-34). In the special case of $D_1(\tau)$ constant, the results
presented in the literature [3,7] are recovered.

\section{Non Linear Porous Media Equation}

For the solutions (27) considered above, the following results
follows readily:
\begin{equation}
 v p(\tau,v)=\frac{N_q^{q-1}}{2(q-2)} \beta(\tau)^{(q-3)/2}
  \frac{\partial}{\partial v}[p(\tau,v)]^{2-q} \ \ .
\end{equation}
Substituting the relationship above in eq.(30) it follows:
\begin{equation}
\frac{\partial p(\tau,v)}{\partial \tau}= D_2(\tau)
\frac{\partial^2}{\partial v^2} [p(\tau,v)]^{2-q} \ \ ,
\end{equation}
where
\begin{equation}
 D_2(\tau)=\frac{N_q^{q-1}}{q-2} \beta(\tau)^{(q-3)/2}
 \left[\frac{1}{2}-\beta(\tau)D(\tau) \right] \ \ .
\end{equation}
As a consequence, eq. (36) is the well known  non-linear porous
media equation, widely used in condensed matter physics and
considered recently in ref. [8]. The current of particles is given
by:
\begin{equation}
 j(\tau,v)=-D_2(\tau)\frac{\partial}{\partial v}[p(\tau,v)]^{2-q} \ \
 ,
\end{equation}
which generalizes the Fick law (indeed, for $q=1$ the classic Fick
law is recovered).

Clearly, the time dependent  Tsallis distribution (18) is solution
of eq. (36) as well. $Z_q(\tau)$ and $\beta(\tau)$ are connected
by eq. (19) and the solution acquires the form (27) while the
evolution law for $\beta(\tau)$ follows from $D_2(\tau)$. The
final differential equation is
\begin{equation}
\frac{d \beta(\tau)}{d\tau}= 4(q-2)N_q^{1-q}D_2(\tau)
\beta(\tau)^{(5-q)/2} \ \ ,
\end{equation}
that is solved readily to obtain
\begin{eqnarray}
\beta(\tau)=\beta(0)\Bigg[1+2(q-2)(q-3)N_q^{1-q}&& \nonumber \\
\times\beta(0)^{(3-q)/2}\int_0^\tau
D_2(\tau)d\tau\Bigg]^{2/(q-3)}&& \ \ .
\end{eqnarray}
From eq. (37) and (39) it follows that $D_2(\infty)=0$, a
condition that guarantees that the current (38) vanishes for
$\tau\rightarrow\infty$, as needed to obtain an equilibrium state.

\section{Conclusions}
In the present work, the same time dependent Tsallis statistical
distribution given in eq. (27) is solution of all three equations
considered, eq. (17), eq.(30) and eq. (36) and describes anomalous
diffusion. The time evolution of $\beta(\tau)$ is the same in all
the cases considered and can be expressed in terms of the three
functions $D(\tau)$, $D_1(\tau)$, $D_2(\tau)$ that are
inter-related.

The relationships derived above allow us to interpret the states
(27) at a microscopic level. Such states are solutions of non
linear FP equations  (eqs.(30) and (36)) and consequently describe
anomalous diffusion (when $q\neq1$). But the results obtained
above, prove that anomalous diffusion can also be described with
the linear FP equation (17), which has a variable diffusion
coefficient given by eq. (16).

The primary results of this equivalence is that the linear FP
equation (17) can be related directly to the microscopic dynamic
model expressed by the Langevin equation (1) linking in this way
macroscopic processes described by anomalous diffusion with
microscopic processes characterized by multiplicative noise.

\end{document}